\documentclass[aip,reprint,amsmath,amssymb,floatfix]{revtex4-2}

\usepackage[acronym,symbols,nogroupskip,nomain,nonumberlist,nopostdot,toc]{glossaries}
\usepackage{siunitx}
\usepackage[version=4]{mhchem}
\usepackage{booktabs}
\usepackage{graphicx}
\usepackage{xcolor}
\usepackage{amsmath}
\usepackage{braket}
\usepackage{hyperref}
\usepackage[capitalize]{cleveref} 
\usepackage[subpreambles=false]{standalone}
\usepackage{tikz}
\usepackage{xcolor}

\usetikzlibrary{arrows}
\usetikzlibrary{shapes.geometric}

\definecolor{purple}{HTML}{6928c3}
\definecolor{orange}{HTML}{f17400}
\definecolor{light-gray}{gray}{0.66}

\newcommand{\mbf}[1]{\mathbf{#1}}

\DeclareSIUnit\angstrom{\text{Å}}
\DeclareSIUnit\bohr{\text{\ensuremath{a}}_{0}}
\DeclareSIUnit\hartree{\text{\ensuremath{E}}_{\mathrm{h}}}

\newacronym{adapt}{ADAPT-VQE}{adaptive derivative-assembled pseudo-Trotter ansatz VQE}
\newacronym{qadapt}{q-ADAPT}{qubit-ADAPT-VQE}
\newacronym{as}{AS}{active space}
\newacronym{api}{API}{application programming interface}

\newacronym{cbm}{CBM}{conduction band minimum}
\newacronym{ccsd}{CCSD}{coupled cluster singles and doubles}
\newacronym{csf}{CSF}{configuration state function}

\newacronym{dft}{DFT}{density functional theory}

\newacronym{eri}{ERI}{electron-repulsion integral}

\newacronym{fci}{FCI}{full configuration interaction}

\newacronym{gga}{GGA}{generalized gradient approximation}
\newacronym{gpw}{GPW}{Gaussian and plane waves}

\newacronym[longplural=hardware efficient ansatzes]{hea}{HEA}{hardware efficient ansatz}
\newacronym{hf}{HF}{Hartree--Fock}
\newacronym{hpc}{HPC}{high-performance computing}

\newacronym{io}{I/O}{input/output}
\newacronym{ipc}{IPC}{inter-process communication}

\newacronym{jw}{JW}{Jordan--Wigner}

\newacronym{ks}{KS}{Kohn--Sham}
\newacronym{ksdft}{KS-DFT}{Kohn--Sham density functional theory}

\newacronym{lda}{LDA}{local density approximation}
\newacronym{lr}{LR}{long-range}

\newacronym{mo}{MO}{molecular orbital}

\newacronym{pbe}{PBE}{Perdew, Burke, and Ernzerhof}
\newacronym{pl}{PL}{photoluminescence}

\newacronym{qeom}{qEOM}{quantum equation of motion}
\newacronym{qpe}{QPE}{quantum phase estimation}

\newacronym[longplural=reduced density matrices]{rdm}{RDM}{reduced density matrix}
\newacronym{1rdm}{1-RDM}{one-particle RDM}
\newacronym{2rdm}{2-RDM}{two-particle RDM}
\newacronym{rs}{RS}{range-separation}
\newacronym{rsdft}{rsDFT}{range-separated density functional theory}

\newacronym{scf}{SCF}{self-consistent field}
\newacronym{sd}{SD}{Slater determinant}
\newacronym{sdk}{SDK}{software development kit}
\newacronym{si}{SI}{Supplementary Information}
\newacronym{sr}{SR}{short-range}

\newacronym{tddft}{TDDFT}{time-dependent density functional theory}
\newacronym{tdl}{TDL}{thermodynamic limit}

\newacronym{ucc}{UCC}{unitary coupled cluster}
\newacronym{uccsd}{q-UCCSD}{quantum UCC singles and doubles}
\newacronym{uml}{UML}{unified modeling language}

\newacronym{vcc}{VCC}{variational coupled cluster}
\newacronym{vqe}{VQE}{variational quantum eigensolver}

\newacronym{wf}{WF}{wave function}

\newacronym{xcf}{XCF}{exchange-correlation functional}

\begin{document}

\title{A general framework for active space embedding methods: applications in quantum computing}

\author{Stefano Battaglia}
\email{stefano.battaglia@chem.uzh.ch}
\thanks{These authors contributed equally to this work.}
\affiliation{Department of Chemistry,
    University of Zurich,
    Winterthurerstrasse 190,
    CH-8057 Zürich,
    Switzerland
}

\author{Max Rossmannek}
\email{oss@zurich.ibm.com}
\thanks{These authors contributed equally to this work.}
\affiliation{Department of Chemistry,
    University of Zurich,
    Winterthurerstrasse 190,
    CH-8057 Zürich,
    Switzerland
}
\affiliation{IBM Quantum, IBM Research – Zurich, CH-8803 Rüschlikon, Switzerland}

\author{Vladimir V. Rybkin}
\altaffiliation[now at: ]{HQS Quantum Simulations GmbH,
    Rintheimer Strasse 23,
    DE-76131 Karlsruhe,
    Germany
}
\affiliation{Department of Chemistry,
    University of Zurich,
    Winterthurerstrasse 190,
    CH-8057 Zürich,
    Switzerland
}

\author{Ivano Tavernelli}
\affiliation{IBM Quantum, IBM Research – Zurich, CH-8803 Rüschlikon, Switzerland}

\author{Jürg Hutter}
\affiliation{Department of Chemistry,
    University of Zurich,
    Winterthurerstrasse 190,
    CH-8057 Zürich,
    Switzerland
}

\begin{abstract}
    We developed a general framework for hybrid quantum-classical computing of molecular and periodic
    embedding approaches based on an orbital space separation of the fragment and environment degrees of
    freedom. 
    We demonstrate its potential by presenting a specific implementation of periodic range-separated DFT coupled
    to a quantum circuit ansatz, whereby the variational quantum eigensolver and the quantum equation-of-motion
    algorithm are used to obtain the low-lying spectrum of the embedded fragment Hamiltonian.
    Application of this scheme to study localized
    electronic states in materials is showcased through the accurate prediction of the optical properties of
    the neutral oxygen vacancy in magnesium oxide (MgO). Despite some discrepancies in the position of the main
    absorption band, the method demonstrates competitive performance compared to state-of-the-art \textit{ab initio}
    approaches, particularly evidenced by the excellent agreement with the experimental photoluminescence
    emission peak.
\end{abstract}

\date{\today}

\maketitle

\section{Introduction}
\label{sec:intro}

Over the past decades, the electronic structure simulation of molecular and solid-state systems
has assumed an increasingly important role in the research and development of new materials.
While the exact computation of ground and excited state properties poses an exponentially scaling
problem through the solution of the Schr{\"o}dinger equation, many methods have been developed to various degrees
of approximation rendering their implementation feasible.
In particular, \gls{dft} has established itself as a cheap yet effective mean for the simulation
of a wide range of systems of interest.
However, due to its nature, it falls short in the description of problems that contain strongly
correlated electrons.
In general, the accurate description of such systems requires methods that consider multiple
\glspl{csf} in order to capture the complex entanglement between the electrons.
These methods of greater accuracy come at increasingly higher computational costs approaching
the theoretical exponential scaling of the exact solution, therefore limiting their applicability
to problems of relatively small sizes.

In the past decade, significant progress has been made in the development of quantum computers,
which provide access to a new computational paradigm that promises to overcome this exponential
barrier~\cite{Barkoutsos2018_ParticleHole, motta_rice_2022}.
This feat can be achieved by encoding the exponentially scaling \gls{wf} in a polynomial number
of \emph{qubits}, the fundamental processing units of a quantum computer, which
enables the efficient representation of an otherwise inaccessible computational space.
Many algorithms have been developed to leverage this representation for the computation of ground
and excited-state properties of chemical systems~\cite{Alexeev2023}.
However, these algorithms exceed the quantum computational capabilities of state-of-the-art devices
when applied to large problems.
On one hand, this is due to the deep quantum circuits that arise when encoding a fermionic \gls{wf}
in a set of qubits, while ensuring the physical nature of the generated superposition.
The long runtime of such quantum circuits exceeds the decoherence times of the available hardware,
resulting in the accumulation of errors which require \emph{error correction} to be feasible before
we can gain any significant results~\cite{Alexeev2023}.
On the other hand, algorithms which trade the execution of a single long quantum circuit for a
(variational) optimization problem involving many shorter circuits quickly become
infeasible; in particular when targeting systems that are beyond the current capabilities of
classical computational hardware~\cite{Peruzzo2014}.

As a short-term solution to these limitations, combined with the interest of the computational
chemistry and materials science communities to leverage this new computing platform, we have seen in
recent years increasing efforts in the development of new hybrid quantum-classical algorithms. 
In particular, many embedding methods have been adapted or newly developed to leverage the quantum
resources for the treatment of embedded fragments, where in many cases the computation of a small region
of a larger system is offloaded to the quantum computer, while the rest is carried out on classical hardware~\cite{Rossmannek2021,Huang2022,Li2022a,Otten2022,Vorwerk2022,Huang2023a,Izsak2023,Liu2023,Rossmannek2023}.
These efforts have been very important for exploring the current practical limitations of noisy devices
and for benchmarking quantum computational approaches for the simulation of chemical and solid-state
systems.
Crucially, embedding methods also provide the means to systematically scale the problems of interest
alongside the development of quantum computing hardware, such that the infrastructure and resources
invested now will eventually allow to reach system sizes and an accuracy beyond what is currently feasible.

Adding to these efforts, this work presents a general framework for the implementation of
active space embedding methods to couple classical and quantum computational codes.
In particular, building up on an earlier work by some of the authors~\cite{Rossmannek2021}, we
extend the range-separated \gls{dft} embedding scheme to allow embedding into periodic environments.
The new and generally improved implementation, coupling the CP2K package~\cite{Kuhne2020}
and Qiskit Nature~\cite{Qiskit,QiskitNature}, is capable of simulating both molecular and periodic systems alike.
The communication between the two codes is handled through \emph{message passing} permitting future
extensions as well as providing a scalable path towards quantum-centric supercomputing.

The rest of this paper is structured as follows.
In Section \ref{sec:theory} we recap the theory of active space embedding routines and extend
the previously published range-separated \gls{dft} embedding to periodic systems.
We also review the most relevant concepts of quantum computing required for the simulation of
electronic structure systems.
Section \ref{sec:implementation} discusses the details of the integration between CP2K and Qiskit Nature.
We think that this work will present useful and generally applicable guidelines for other hybrid
quantum-classical integrations in the future.
In Section \ref{sec:results} we present and discuss the results obtained by applying the developed
methodology to study the optimal properties of the neutral oxygen vacancy in magnesium oxide.
Finally, Section \ref{sec:conclusion} concludes this work.

\section{Theory}
\label{sec:theory}

We present a general framework for active space embedding approaches,
where a subset of electrons and orbitals of a system --- \emph{the fragment} --- are embedded in
an effective potential generated by the remaining electrons of the systems and all the ion
cores --- \emph{the environment}.
We do this in the framework of multiconfigurational \gls{rsdft}~\cite{Savin1996,Leininger1997,Pollet2002,Toulouse2005,Fromager2007,Hedegard2018,Pernal2022}
and the \gls{gpw} approach~\cite{Lippert1997}, whereby the embedding potential is obtained from
a mean-field method, while the fragment Hamiltonian is solved with a correlated wave function ansatz.
The approach and infrastructure we have developed is completely general; it can treat both
molecular and periodic systems, it supports spin-polarized and unpolarized calculations,
it describes the core electrons explicitly or through pseuopotentials, and
can be combined with both classical \gls{wf} and quantum circuit ansatzes alike.

\subsection{Active space embedding}
\begin{figure}
    \centering
    \includegraphics[width=\linewidth]{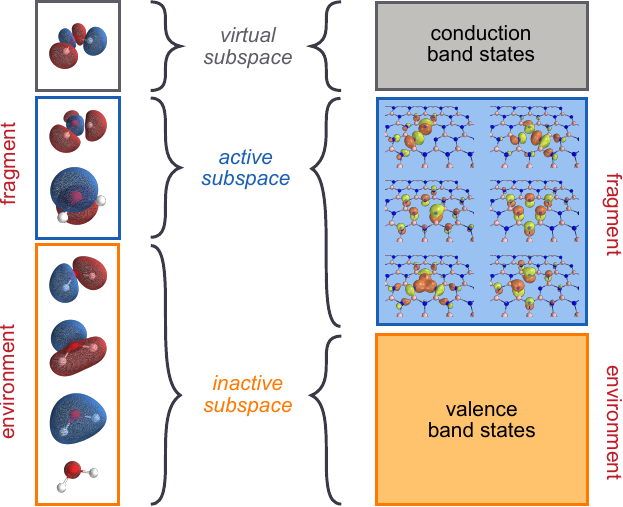}
    \caption{Example of possible selections of active and inactive spaces for the water
    molecule (left) and for a positively charged boron vacancy in hexagonal boron nitride (right).
    In both cases, only a small number of orbitals is included in the active space: for water,
    it is spanned by the HOMO-LUMO pair, while for boron nitride, by the localized defect orbitals.
    Notice that also the number of active electrons must be specified, though, once the active
    orbitals have been selected, the number of active electrons is easily identified.}
    \label{fig:orb_space}
\end{figure}

To introduce the framework for active space embedding methods, we start with
the second-quantized electronic Hamiltonian in the Born-Oppenheimer approximation.
This can be written in atomic units as
\begin{equation}
    \label{eq:H_tot}
    \hat{H} = \sum_{pq} h_{pq} \hat{a}_p^\dagger \hat{a}_q +
    \frac{1}{2} \sum_{pqrs} g_{pqrs} \hat{a}_p^\dagger\hat{a}_r^\dagger\hat{a}_s\hat{a}_q
    + \hat{V}_{nn}
    \, ,
\end{equation}
where $\hat{V}_{nn}$ is the Coulomb repulsion between the nuclei, while $h_{pq}$ and $g_{pqrs}$
are one- and two-electron integrals given by
\begin{align}
    \label{eq:h_pq}
    h_{pq} &= \Braket{\psi_p(\mbf{x}) | \hat{h} |\psi_q(\mbf{x})} \, , \\
    \label{eq:g_pqrs}
    g_{pqrs} &= \Braket{\psi_p(\mbf{x}) \psi_r(\mbf{x}') | \hat{g} | \psi_q(\mbf{x}) \psi_s(\mbf{x}')} \, .
\end{align}
The variable $\mbf{x}$ ($\mbf{x}'$) is a compound variable for the electron coordinates in space,
$\mbf{r}$ ($\mbf{r}'$), and its spin degree of freedom.
The operators $\hat{h}$ and $\hat{g}$ in \cref{eq:h_pq,eq:g_pqrs} account for the kinetic energy of the electrons,
the electron-nuclear attraction, and the electron-electron repulsion, and are defined as
\begin{align}
    \hat{h}(\mbf{r}) &= -\frac{1}{2} \nabla^2 + \sum_P \frac{-Z_P}{\lvert \mbf{r} - \mbf{R}_P \rvert} \, , \\
    \hat{g}(\mbf{r}, \mbf{r}') &= \frac{1}{|\mbf{r} - \mbf{r}'|} \, ,
\end{align}
where $P$ labels the ion cores, while $Z_P$ and $\mbf{R}_P$ denote the corresponding nuclear charges
and nuclear positions, respectively.
The indices $p, q, r, s$ label general one-particle functions (spin-orbitals), whose corresponding sums
appearing in \cref{eq:H_tot} run through the entire basis set.
The operator $\hat{a}^\dagger_{p}$ ($\hat{a}_{p}$) is a creation (annihilation) operator adding
(removing) an electron to (from) spin-orbital $\psi_p(\mbf{x})$.
In an embedding approach like the one presented here, the fragment of interest is defined
in terms of an \emph{active space} consisting of a handful of \emph{active electrons} and \emph{active orbitals}.
All the electrons that are not part of the active space typically occupy the low-energy states
of the system and are called \emph{inactive orbitals}; see \cref{fig:orb_space} for an example active space
selection in molecules and materials.
Once the active space has been identified, a corresponding embedded fragment Hamiltonian can be defined
in a manner completely analogous to \cref{eq:H_tot}, that is
\begin{equation}
    \label{eq:H_frag}
    \hat{H}^{\text{frag}} = \sum_{uv} V^\text{emb}_{uv} \hat{a}^\dagger_u \hat{a}_v +
    \frac{1}{2} \sum_{uvxy} g_{uvxy} \hat{a}^\dagger_u \hat{a}^\dagger_x \hat{a}_y \hat{a}_v \, ,
\end{equation}
with the only difference that the sums are limited to the active orbitals, labeled by the indices $u,v,x,y$,
and that the one-electron integrals, $h_{pq}$, have been replaced by the elements of an embedding potential,
$V^\text{emb}_{uv}$. This potential accounts for the interactions between the inactive and active electrons
in addition to the contributions from the one-electron integrals.
Notice that until this point this formulation is completely general; we have not yet specified how to compute
the embedding potential. In principle, one could define an operator that explicitly accounts for
all many-body interactions between the inactive and active subsystems, though, probably this would result in a
methodology that is computationally as expensive as solving directly for the entire problem with such an
approach.
Therefore, in practice, the embedding potential introduces approximations to describe the low-energy degrees
of freedom and the interactions between active and inactive subsystems. For example, one such option would be
to use the \gls{hf} approximation for the inactive electrons, such that the active electrons only interact
with the active ones in a mean-field manner. In this case, $V^\text{emb}_{uv}$ would simply correspond to the elements of
the Fock matrix. Similarly, as we will discuss in more detail in the next section, describing the environment
using \gls{dft} translates into an embedding potential similar to the \gls{ks} one.
It is important to realize that in general, $V^\text{emb}_{uv}$ always depends on the inactive electronic
degrees of freedom, but possibly also on the active subsystem, in which case the resulting embedding scheme
has to be solved self-consistently (see \cref{fig:cp2k-qiskit}).

To compute the total energy of the system, we can start from the expectation value of \cref{eq:H_tot} with
respect to the total \gls{wf} of the system, that is
\begin{equation}
    E = \Braket{\Psi|\hat{H}|\Psi} = \sum_{pq} h_{pq}D_{pq}
      + \frac{1}{2} \sum_{pqrs} g_{pqrs} d_{pqrs}
      + V_{nn} \, ,
\end{equation}
where
\begin{align}
    D_{pq}   &= \braket{\Psi|\hat{a}_p^\dagger \hat{a}_q | \Psi} \, , \\
    d_{pqrs} &= \braket{\Psi|\hat{a}^\dagger_p \hat{a}^\dagger_r
                             \hat{a}_s         \hat{a}_q | \Psi} \, ,
\end{align}
are the elements of the one- and two-particle \glspl{rdm}, $\mbf{D}$ and $\mbf{d}$, respectively.
By separating the \gls{1rdm} into inactive and active components, $\mbf{D} = \mbf{D}^I + \mbf{D}^A$,
and factorizing the elements of the inactive \gls{2rdm} into a product of \glspl{1rdm}
(see Appendix A.1 and A.2 of~\citet{Rossmannek2021} for a detailed derivation),
we can express the total energy as a sum of inactive and active parts, $E = E^I + E^A$, with 
\begin{align}
    \label{eq:E_I}
    E^I = \sum_{ij} (h_{ij} + V^\text{emb}_{ij}) D^I_{ij}
        + V_{nn} \, ,
\end{align}
and
\begin{align}
    \label{eq:E_A}
    E^A = \sum_{uv} V^\text{emb}_{uv} D^A_{uv}
    + \frac{1}{2} \sum_{uvxy} g_{uvxy} d^A_{uvxy} \, . 
\end{align}
In \cref{eq:E_I,eq:E_A}, the indices $i,j$ label inactive orbitals and the superscripts $I$ and $A$
on the density matrix elements emphasize to which subspace they belong (even though the indices and
sums implicitly account for that information).
At last, notice that the choice of one-particle functions is completely general: one can choose
localized molecular orbitals in case of molecules, crystalline orbitals or Wannier functions in 
solid-state systems, or a combination thereof, \textit{e.g.} to describe point defects in
materials.

\subsection{Periodic range-separated DFT embedding}\label{sec:rsdft}
As a starting point for introducing the periodic range-separated \gls{dft} embedding,
we can rely on the working equations of~\citet{Rossmannek2021} that we will
briefly recap in the following.
The first ingredient is the definition of the one-particle embedding potential, with elements
\begin{align}
    V^{\text{emb}}_{pq}
    &= F^{I,\text{LR}}_{pq}
    + V^{\text{SR}}_{H,pq}[\rho^I]
    + V^{\text{SR}}_{H,pq}[\rho^A]
    + V^{\text{SR}}_{xc,pq}[\rho] \, ,
    \label{eq:vemb_rsdft}
\end{align}
where the elements of the inactive long-range Fock operator are defined as
\begin{align}
    F^{I,\text{LR}}_{pq}
    &= h_{pq}
    + V^{\text{LR}}_{H,pq}[\rho^I]
    + V^{\text{LR}}_{\text{HFX},pq}[\rho^I] \, ,
    \label{eq:fock_inact_lr}
\end{align}
along with the classical Hartree potential, $V_{H}[\rho]$, the exact Hartree-Fock
exchange potential, $V_{\text{HFX}}[\rho]$, and the DFT exchange-correlation potential,
$V_{xc}[\rho]$, evaluated over the indicated electron densities, $\rho^I, \rho^A$ and
$\rho = \rho^I + \rho^A$ (see the \gls{si} for the explicit definition
of these operators in a one-particle basis).
The two-electron integrals over the Coulomb operator are split into  \gls{lr} and \gls{sr}
components,
\begin{align}
    \hat{g} &= \hat{g}^{\omega,\text{LR}} + \hat{g}^{\omega,\text{SR}} \, \\
    &= \frac{\text{erf} (\omega |\mbf{r} - \mbf{r}'|)}{|\mbf{r} - \mbf{r}'|}
    +  \frac{\text{erfc}(\omega |\mbf{r} - \mbf{r}'|)}{|\mbf{r} - \mbf{r}'|}
    \label{eq:g_rs}
\end{align}
which give rise to the superscripts $\text{LR}$ and $\text{SR}$ in
\cref{eq:vemb_rsdft,eq:fock_inact_lr}.
The range separation is obtained with the error function and its complement (as indicated
by \cref{eq:g_rs}), where $\omega$ is the \gls{rs} parameter of units \unit{\bohr^{-1}}.

In practice, two issues arise for the direct computation of the inactive energy and
embedding potential according to \cref{eq:E_I,eq:vemb_rsdft}.
First, it is computationally disadvantageous when the inactive subsystem becomes very large.
Second, for periodic calculations, the sums concerning the electron-electron, electron-nuclear,
and nuclear-nuclear interactions are conditionally convergent, and cannot be easily separated
into inactive and active components.
Hence, we express the inactive terms indirectly as the difference between the total system and
the (localized) active subsystem. We can achieve this by defining the inactive \gls{1rdm} and
electron density as $\mathbf{D}^I = \mathbf{D} - \mathbf{D}^A$ and $\rho^I = \rho - \rho^A$,
respectively.
Reformulating \cref{eq:fock_inact_lr} by replacing $\rho^I = \rho - \rho^A$, yields
\begin{align}
    F^{I,\text{LR}}_{pq}
    = F^{\text{tot}}_{pq}
    &- V^{\text{SR}}_{H,pq}[\rho] - V^{\text{SR}}_{xc,pq}[\rho] \nonumber \\
    &- V^{\text{LR}}_{H,pq}[\rho^A] - V^{\text{LR}}_{\text{HFX},pq}[\rho^A] \, ,
    \label{eq:fock_inact_lr_indirect}
\end{align}
where the total \gls{rsdft} Fock operator is defined as
\begin{align}
    F^{\text{tot}}_{pq}
    &= h_{pq} + V_{H,pq}[\rho]
    + V^{\text{LR}}_{\text{HFX},pq}[\rho]
    + V^{\text{SR}}_{xc,pq}[\rho] \, .
    \label{eq:fock_tot}
\end{align}
Inserting the same relation for the inactive electron density as well as
\cref{eq:fock_inact_lr_indirect} into \cref{eq:vemb_rsdft} results in
\begin{align}
    V^{\text{emb}}_{pq}
    &= F^{\text{tot}}_{pq}
    - V^{\text{LR}}_{H,pq}[\rho^A]
    - V^{\text{LR}}_{\text{HFX},pq}[\rho^A] \, .
    \label{eq:vemb_indirect}
\end{align}
%
We can proceed analogously for the expression of the inactive energy, obtaining
\begin{align}
    E^{I} &= E^{\text{tot}}
    - \sum_{uv} F^{\text{tot}}_{uv} D^A_{uv}
    + E^{\text{LR}}_{H}[\rho^A]
    + E^{\text{LR}}_{\text{HFX}}[\rho^A] \, .
    \label{eq:e_inact_indirect}
\end{align}
The active energy component that is needed to compute the total energy, $E = E^I + E^A$, simply
corresponds to the ground state of the fragment Hamiltonian, \cref{eq:H_frag}. Owing to the similar
structure of \cref{eq:H_tot,eq:H_frag}, essentially any electronic structure method can be used in
combination with our embedding scheme. In practice, because the space spanned by the active orbitals
and electrons is relatively small, exact diagonalization or a good approximation thereof is the
method of choice. Electronically excited states can also be targeted by the embedding method,
either by directly calculating the spectrum of $\hat{H}^\text{frag}$ or by linear response. However,
one has to be careful that the inactive subspace is normally optimized for the ground state,
unless some form of state-averaging or orbital optimization similar to classical multiconfigurational
quantum chemical methods is introduced~\cite{Roos1980,Werner1981}.
As will be discussed in the next section, we have used quantum circuit ansatzes
to obtain the ground and excited states energies of \cref{eq:H_frag}.
Owing to the dependence of the embedding potential to the active space electron density, $\rho^A$,
the method chosen to get the spectrum of $\hat{H}^\text{frag}$ should also provide this quantity
(more generally, it should provide the \gls{1rdm}). Crucially, this dependence of $V^\text{emb}$ on
$\rho^A$ implies that our embedding approach requires a self-consistent solution, whereby an updated
active density is obtained at each iteration, which is used to build a refined embedding potential
and updated inactive energy that account for the feedback of the active subsystem on the environment
degrees of freedom. A scheme depicting this self-consistent loop is shown in \cref{fig:cp2k-qiskit},
when discussing the implementation details in Section \ref{sec:implementation}.

Finally, it should be emphasized that \cref{eq:vemb_indirect,eq:e_inact_indirect} generalize
to the molecular and periodic embedding settings, since only the computation of the total Fock
operator and energy are affected by this change, at least when invoking the $\Gamma$ point
approximation.
Furthermore, we point out the limiting cases provided by the \gls{rs} scheme: they allow us
to recover the common \gls{hf} embedding scheme as $\omega$ approaches infinity as well as
\gls{ks} \gls{dft} as $\omega$ approaches zero.
This can be seen in \cref{eq:vemb_indirect,eq:e_inact_indirect}, where the standard
\gls{ks} case is evident as all \gls{lr} terms simply disappear.
The common \gls{hf} embedding is also evident from \cref{eq:fock_tot}, in which the only
\gls{dft}-specific term for the exchange-correlation interaction, $V_{xc}^{\text{SR}}$, vanishes.

\subsection{Quantum computing}

We leverage quantum computing for finding the ground and excited state solutions of
the embedded fragment Hamiltonian (cf.\@\cref{eq:H_frag}).
We do so through the means of the newly developed integration between CP2K~\cite{Kuhne2020} and
Qiskit Nature~\cite{Qiskit,QiskitNature} which we discuss in more detail in Section \ref{sec:implementation}.
In recent years, quantum computing has made significant progress in emerging as a new computational
paradigm that promises great advances for the simulation of chemistry and material science
problems~\cite{Alexeev2023}.
In particular for the latter, the ability to treat localized fragments embedded into periodic systems
using quantum computing platforms is particularly appealing.
While we rely on state-of-the-art hybrid quantum-classical algorithms such as the \gls{vqe}~\cite{Peruzzo2014} and \gls{qeom}~\cite{Ollitrault2020}, the presented embedding framework 
is not coupled to these choices and can leverage any advancements in the field of quantum computing
that are yet to come.
This also holds for the modular integration of the two computational codes as we will show later.
In this section, we briefly review the theoretical foundations of the quantum computational
tools used throughout this work.

\subsubsection{The fermion-to-qubit mapping}
In order to simulate a fermionic system on a quantum computer one must first map the second-quantized
Hamiltonian (cf.\@~\cref{eq:H_frag}) into a form that the quantum computer can work with.
Since the fundamental operational unit of a quantum computer is a two-level system, a so-called
\emph{qubit}, the translation routines are referred to as fermion-to-qubit mappings.
Many such mappings exist~\cite{Jordan1928,Bravyi2002,Bravyi2017,Jiang2020,Shee2022,Miller2023}, the
most common of which is arguably the Jordan-Wigner mapping~\cite{Jordan1928}.
It maps the fermionic creation, $\hat{a}^\dagger_p$, and annihilation, $\hat{a}_p$, operators acting
on spin orbital, $p$, to a tensor product of identities, $\mathcal{I}$, and Pauli matrices,
$\{\hat{\sigma}^X_p,\hat{\sigma}^Y_p,\hat{\sigma}^Z_p\}$, which correspond to the principal
single-qubit rotations along the Cartesian axes.
The mapping can be written as
\begin{align}
    \hat{a}_p &\rightarrow \left( \bigotimes_{q=1}^{p-1} \hat{\sigma}^Z_q \right)
    \hat{\sigma}^-_p \left( \bigotimes_{q=p+1}^{M} \mathcal{I}_q \right) \, , \\
    \hat{a}^\dagger_p &\rightarrow \left( \bigotimes_{q=1}^{p-1} \hat{\sigma}^Z_q \right)
    \hat{\sigma}^+_p \left( \bigotimes_{q=p+1}^{M} \mathcal{I}_q \right) \, ,
\end{align}
where $\hat{\sigma}^\pm = (\hat{\sigma}^X \pm i\hat{\sigma}^Y)/2$ are the so-called \emph{ladder} operators.
This mapping is the most straightforward since it encodes the occupation of one spin orbital into the
occupation of a single qubit.
Therefore, it will encode a fermionic Hamiltonian acting on $M$ spin orbitals into a qubit Hamiltonian
acting on $M$ qubits.
The anti-commutation relations of the fermionic operators are preserved by the chain of $\hat{\sigma}^Z$
rotations on lower-indexed qubits.
The resulting qubit Hamiltonian is a weighted sum of the form~\cite{Barkoutsos2018_ParticleHole}
\begin{align}
    \label{eq:hamil_qubit}
    \hat{H}^{\text{qu}} = \sum_p c_p \hat{P}_p \, ,
\end{align}
where each Pauli string, $\hat{P}_p$, is a tensor product of identities and Pauli matrices.
Crucially, the number of unique terms in $\hat{H}^{\text{qu}}$ scales as $\mathcal{O}(M^4)$, just like
the number of two-body interactions in the original second-quantized Hamiltonian, \cref{eq:H_frag}.

Another very common mapping is the parity mapping~\cite{Bravyi2017} which can be thought of as the \emph{dual} to
the Jordan-Wigner mapping. It encodes the parity information locally on the qubits and delocalizes the
occupation information of the fermionic spin orbitals. This has an added benefit that for particle-number
conserving Hamiltonians, two qubits carry redundant information and may be omitted (which is referred to
as \emph{two-qubit reduction}~\cite{Bravyi2017}). This is the mapping which was used for all simulations in this
work, but any other fermion-to-qubit mapping could be used in its place.

\subsubsection{The variational quantum eigensolver}
The \gls{vqe}~\cite{Peruzzo2014} is a hybrid quantum-classical algorithm to find the ground state
of any Hamiltonian.
It has gained a widespread interest as an alternative to \gls{qpe} since it is more amenable to
near-term quantum computers.
It does so by replacing the execution of a single long-running quantum circuit with the sampling
of many shorter duration circuits.
The fundamental principle of the \gls{vqe} is based on the expectation value computation of an observable,
$\hat{O}$, with respect to some reference state, $\Psi$, as
\begin{align}
    \braket{\hat{O}} = \frac{\braket{\Psi|\hat{O}|\Psi}}{\braket{\Psi|\Psi}} \, .
\end{align}
Finding the ground state of a Hamiltonian, $\hat{H}$, amounts to using a parameterized ansatz
for the \gls{wf}, $\Psi(\boldsymbol{\theta})$, and variationally optimizing the parameters,
$\boldsymbol{\theta}$, with respect to the expectation value, $\braket{\hat{H}}$.
In recent years, many variants of the \gls{vqe} algorithm have been developed (see for instance~\cite{Grimsley2019,Tang2021,Zhang2021a,Anastasiou2024}),
which oftentimes leverage specific structures found in particular ansatzes for the \gls{wf}.
Since we are only simulating the execution of quantum circuits on classical computers,
we do not require improved circuit depths or other benefits brought about by these algorithm variants.
Thus, we employ the unaltered \gls{vqe} algorithm.

\subsubsection{The wave function ansatz}
Choosing an ansatz for the \gls{vqe} algorithm is a difficult task.
On one hand, it has to be expressive enough to contain the true ground state in its parameterized
subspace of the entire Hilbert space.
On the other hand, it should be limited in its circuit depth implementation to ensure that it can be
executed on near-term quantum computers.
Both of these properties can be achieved by means of \glspl{hea}, which can even be tailored to respect
limited qubit connectivity of the quantum computing hardware.
However, optimization of such ansatzes can pose to be challenging due to the large number of variational
parameters~\cite{McClean2018}.
Moreover, chemical problems, such as the ones we are interested in here, are constrained
to the Fock space 
that often exhibit additional symmetries, further restricting the size of the subspace containing
the true physical ground state.
Therefore, chemistry-inspired ansatzes have been developed which are designed to explore only this
physical subspace, a famous example of which is the \gls{ucc} ansatz~\cite{Romero2018}.
Its major drawback is the significant circuit depth overhead associated with the implementation of
these constraints.
Nonetheless, it is still orders of magnitude cheaper than an implementation of the \gls{qpe}
algorithm~\cite{Romero2018}.


\subsubsection{Computing excited states}\label{sec:qeom}
Many different (hybrid) quantum algorithms exist for the computation of excited states
~\cite{Higgott2019,Ollitrault2020,Motta2023}; in this work we rely on the \gls{qeom} method~\cite{Ollitrault2020}.
It has the advantage that, once a ground state solution has been found, one only needs to
perform additional measurements on this optimized \gls{wf}.
Other algorithms, however, may require an entire new optimization procedure to be completed
for each targeted excited state~\cite{Higgott2019}.

The fundamental idea of \gls{qeom} relies on a classical computer solving the \emph{equations of motion}
while a quantum computer is used for the measurement of the matrix elements that go into the system
of equations.
When including only single and double excitations in the operator basis, the measurement cost of
these matrix elements scales just like the measurement cost of the Hamiltonian expectation value
with its number of terms $\mathcal{O}(M^4)$.
This can be seen from the expectation values that give rise to the excitation energies that we are after
\begin{align}
    E_{0n} = \frac{\braket{0|[\hat{O}_n, \hat{H}, \hat{O}^\dagger_n]|0}}
                  {\braket{0|[\hat{O}_n, \hat{O}^\dagger_n]|0}} \, ,
\end{align}
where $\ket{0}$ denotes the ground state and $\hat{O}^\dagger_n = \ket{n}\bra{0}$ is the excitation
operator from the ground state to the $n$-th excited state (and $\hat{O}_n$ is the matching
de-excitation operator).
For more details we refer the interested reader to the original paper by \citet{Ollitrault2020}.

Once again, our choice of using \gls{qeom} is not constraining the future applicability
of any other (hybrid) quantum algorithm to be used for the computation of excited states.

\section{Implementation}\label{sec:implementation}
In this section, we present the implementation details of the integration of CP2K~\cite{Kuhne2020}
and Qiskit Nature~\cite{Qiskit,QiskitNature}.
The developments of this work have been released as part of CP2K v2024.1, Qiskit Nature v0.7.0, as well
as a separate module handling more specific parts of the integration called
\texttt{qiskit-nature-cp2k}~\cite{QiskitNatureCP2K}.
In the first part, we discuss the technical aspects and challenges.
Later, we review the future scalability and extensibility of this design.

\subsection{Technical realization}
Interfacing CP2K with Qiskit Nature for the implementation of an iterative embedding scheme
poses a number of challenges.
While CP2K is primarily written in Fortran and provides the means to efficiently run highly
parallelized simulations in a variety of computational setups,
Qiskit Nature (and the underlying Qiskit \gls{sdk}) is mostly developed in Python and has not
yet\footnote{At the time of writing.} reached
a computational maturity comparable to CP2K.
When Qiskit Nature was coupled to other Python-based computational programs in the past, they could easily
share the same Python runtime execution environment and, thus, share all data directly in
memory~\cite{Rossmannek2021,Rossmannek2023}.
For the integration discussed here, this was not possible in such a straightforward manner.
Instead, our implementation relies on a \emph{message passing} protocol in order to exchange 
data between the two codes.
Particularly, for this initial implementation the messages and data are sent over a \emph{socket} file.
This is inspired by a similar architecture used by the i--Pi project~\cite{Kapil2019}.
Additional technical details are available in the \gls{si}.

A \emph{socket} is an \gls{api} used for \gls{ipc}.
Using this protocol, it is possible for the communicating processes to run on the same physical machine
or different ones connected via the internet. The calculation proceeds identically in both scenarios
with the only difference being the latency of the communication. However, this is not of a concern to
us, since the rate-limiting factor of the communication is (in any case) the bandwidth of the connection
to the quantum hardware.
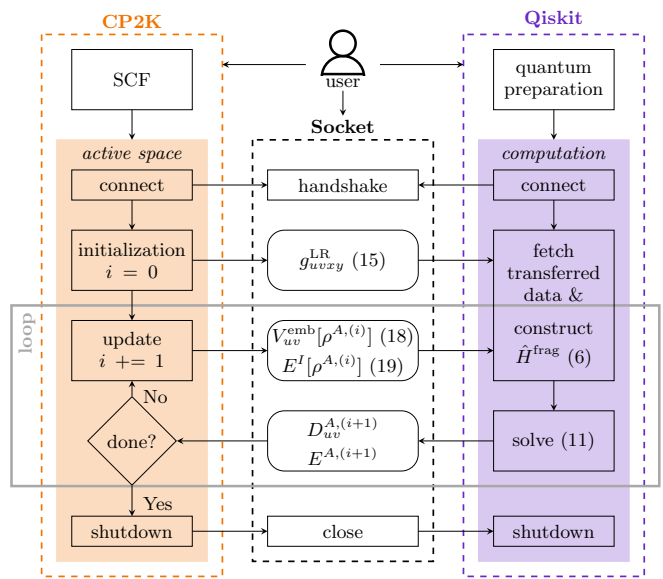
\begin{figure}
    \centering
    \begin{tikzpicture}

\draw (3.5, -1.75) [dashed,thick,draw=black,local bounding box=socket_box] rectangle (6.5, -8.75);
\node at (socket_box.north) [anchor=south] {\textbf{Socket}};

\draw (0.0, 0.0) [dashed,thick,draw=orange,local bounding box=cp2k_box] rectangle (3.0, -9.0);
\node at (cp2k_box.north) [orange,anchor=south] {\textbf{CP2K}};
\draw (0.25, -1.75) [nearly transparent,draw=orange,fill=orange,local bounding box=cp2k_active] rectangle (2.75, -8.75);
\node at (cp2k_active.north) [anchor=north] {\emph{active space}};

\draw (7.0, 0.0) [dashed,thick,draw=purple,local bounding box=qiskit_box] rectangle (10.0, -9.0);
\node at (qiskit_box.north) [purple,anchor=south] {\textbf{Qiskit}};
\draw (7.25, -1.75) [nearly transparent,draw=purple,fill=purple,local bounding box=qiskit_active] rectangle (9.75, -8.75);
\node at (qiskit_active.north) [anchor=north] {\emph{computation}};

\draw (-0.5, -4.5) [very thick,draw=light-gray,local bounding box=loop] rectangle (10.25, -7.5);
\node at (loop.north west) [below left,rotate=90,light-gray] {\textbf{loop}};

\node at (5.0, -0.5) [name=user,text width=1cm,text height=0.5cm] {};
\node at ([yshift=-14pt]user) [anchor=south,align=center] {user};
\begin{scope}
    \clip ([xshift=-12pt,yshift=5pt]user) rectangle ([xshift=12pt,yshift=-5pt]user);
    \draw [very thick] (user) [below=6pt] circle (10pt);
    \draw [very thick] ([xshift=8pt,yshift=-4.5pt]user.west) -- ([xshift=-8pt,yshift=-4.5pt]user.east);
\end{scope}
\draw [very thick] ([yshift=10pt]user) circle (6pt);
\draw [-stealth] (user.west) -- (3.0, -0.5);
\draw [-stealth] (user.east) -- (7.0, -0.5);
\draw [-stealth] ([yshift=-2pt]user.south) -- (5.0, -1.35);

\draw (3.75, -2.25) [local bounding box=socket_handshake] rectangle node [align=center] {handshake} (6.25, -2.75);
\draw (3.75, -8.0) [local bounding box=socket_shutdown] rectangle node [align=center] {close} (6.25, -8.5);
\draw (3.75, -3.25)
    [rounded corners=8,draw=black] rectangle
    node [align=center,text width=2.75cm] {$g^{\text{LR}}_{uvxy}$ (\ref{eq:g_rs})}
    (6.25, -4.25);
\draw (3.75, -4.75)
    [rounded corners=8,draw=black] rectangle
    node [align=center,text width=2.75cm] {$V^{\text{emb}}_{uv}[\rho^{A,(i)}]$ (\ref{eq:vemb_indirect}) \\[3pt] $E^{I}[\rho^{A,(i)}]$ (\ref{eq:e_inact_indirect})}
    (6.25, -5.75);
\draw (3.75, -6.25)
    [rounded corners=8,draw=black] rectangle
    node [align=center,text width=2.75cm] {$D^{A,(i+1)}_{uv}$ \\[3pt] $E^{A,(i+1)}$}
    (6.25, -7.25);

\draw (0.5, -0.25) [draw=black] rectangle node {SCF} (2.5, -1.25);
\draw [-stealth] (1.5, -1.25) -- (1.5, -1.75);
\draw (0.5, -2.25) [local bounding box=cp2k_handshake] rectangle node {connect} (2.5, -2.75);
\draw [-stealth] (cp2k_handshake.east) -- (socket_handshake.west);
\draw [-stealth] (1.5, -2.75) -- (1.5, -3.25);
\draw
    (0.5, -3.25) [draw=black] rectangle
    node [align=center,text width=2.0cm] {initialization \\ $i = 0$}
    (2.5, -4.25);
\draw [-stealth] (1.5, -4.25) -- (1.5, -4.75);
\draw [-stealth] (2.5, -3.75) -- (3.75, -3.75);
\draw
    (0.5, -4.75) [draw=black] rectangle
    node [align=center,text width=2.0cm] {update \\ $i\mathrel{{+}{=}}1$}
    (2.5, -5.75);
\draw [-stealth] (2.5, -5.25) -- (3.75, -5.25);
\draw (1.5, -6.75) node [diamond,draw=black,name=done] {done?};
\draw [stealth-] (done.east) -- (3.75, -6.75);
\draw [-stealth] (done.north) node [right=2pt] {No} -- (1.5, -5.75);
\draw [-stealth] (done.south) node [below right=2pt] {Yes} -- (1.5, -8.0);
\draw (0.5, -8.0) [local bounding box=cp2k_shutdown] rectangle node {shutdown} (2.5, -8.5);
\draw [-stealth] (cp2k_shutdown.east) -- (socket_shutdown.west);

\draw (7.5, -0.25) [draw=black] rectangle node [align=center,text width=2.5cm] {quantum preparation} (9.5, -1.25);
\draw [-stealth] (8.5, -1.25) -- (8.5, -1.75);
\draw (7.5, -2.25) [local bounding box=qiskit_handshake] rectangle node {connect} (9.5, -2.75);
\draw [-stealth] (qiskit_handshake.west) -- (socket_handshake.east);
\draw [-stealth] (8.5, -2.75) -- (8.5, -3.25);
\draw
    (7.5, -3.25) [draw=black] rectangle
    node [dashed,align=center,text width=1.75cm] {fetch transferred data \& \\[5pt] construct \\[3pt] $\hat{H}^{\text{frag}}$ (\ref{eq:H_frag})}
    (9.5, -5.75);
\draw [-stealth] (6.25, -3.75) -- (7.5, -3.75);
\draw [-stealth] (6.25, -5.25) -- (7.5, -5.25);
\draw [-stealth] (8.5, -5.75) -- (8.5, -6.25);
\draw
    (7.5, -6.25) [draw=black] rectangle
    node [dashed,align=center,text width=2.0cm] {solve (\ref{eq:E_A})}
    (9.5, -7.25);
\draw [stealth-] (6.25, -6.75) -- (7.5, -6.75);
\draw (7.5, -8.0) [local bounding box=qiskit_shutdown] rectangle node {shutdown} (9.5, -8.5);
\draw [stealth-] (qiskit_shutdown.west) -- (socket_shutdown.east);

\end{tikzpicture}
    \caption{
        Workflow diagram depicting the interaction of CP2K and Qiskit Nature.
        The user configures the two classical processes and the socket for the \gls{ipc}.
        Each process then follows the computational steps (rectangular boxes) outlined inside
        of their respective frames.
        The data that gets computed and transferred is indicated by the rounded boxes.
        Numbers in parentheses refer to the respective equations in this manuscript.
        The self-consistent embedding requires a loop which is highlighted by the gray box.
        This loop is terminated based on the decision (diamond shape) taken by the CP2K process.
    }
    \label{fig:cp2k-qiskit}
\end{figure}

\cref{fig:cp2k-qiskit} summarizes the computational workflow of our integration between CP2K and Qiskit Nature.
The diagram depicts a user who has to configure three parts of their calculation;
the CP2K and Qiskit (Nature) processes depicted on the left and right, respectively,
as well as the socket itself via which the messages are passed between the two codes.
Both computational codes will start in parallel.
While CP2K starts out by finding a low-level solution to the entire system (SCF),
Qiskit can use this time to perform certain preparational tasks that are unique to the execution on
quantum computing hardware and do not require problem-specific data.
For both programs, the user has full flexibility to leverage their respective capabilities during
these initial steps.
Upon completion of their respective steps, both codes will synchronize by performing a handshake
through the socket.
If either process reaches this point before the other, it awaits the other one.
CP2K reaches this point inside its \emph{active space} module which was released as part of CP2K v2024.1.
The input to this module configures the active fragment to be embedded into its environment and computes
the one- and two-body terms of the active space Hamiltonian (\emph{i.e.} the fragment Hamiltonian).
It allows both single-shot and iterative embedding routines to be performed.
In \cref{fig:cp2k-qiskit} we have depicted the \gls{rsdft} embedding protocol presented in Section \ref{sec:rsdft}.
However, some of the components do not change throughout the course of the self-consistent embedding.
These can be pre-computed only once during the \emph{initialization} procedure.
A key example are the \gls{lr}-\glspl{eri} (cf.\@~\cref{eq:g_pqrs,eq:g_rs}).
Therefore, these only have to be transferred to the Qiskit process once.
Components which depend on the active density of the current iteration, $\rho^{A,(i)}$,
have to be updated and exchanged during every iteration of the loop indicated by the gray frame
in \cref{fig:cp2k-qiskit}.
During every such iteration, Qiskit Nature constructs the Hamiltonian of the active fragment (cf.\@~\cref{eq:H_frag})
using the \gls{lr}-\glspl{eri} and embedding potential, $V^{\text{emb}}_{uv}[\rho^{A,(i)}]$,
(cf.\@~\cref{eq:vemb_indirect}).
It then proceeds with finding the ground-state solution to this Hamiltonian using the quantum
circuit ansatz specified by the user.
Upon completion, it will return the active energy, $E^{A,(i+1)}$, and active \gls{1rdm},
$D^{A,(i+1)}_{uv}$, to the CP2K process.
CP2K will then perform a convergence check based on the total energy, $E = E^I + E^{A,(i+1)}$.
If convergence has not been reached, CP2K and Qiskit will return back to their respective steps
of the embedding protocol to proceed with another iteration.
If this check succeeds, both processes will be signalled to terminate.
During their \emph{shutdown} procedures, both processes may perform additional post-processing steps.
For example, Qiskit Nature may compute additional properties using the final ground-state \gls{wf} including
the computation of excited state energies.

\subsection{Future scalability}
At this point one might wonder how scalable this design is for the future.
Indeed, the transfer of the two-body integrals is the limiting factor here.
If CP2K and Qiskit Nature were able to leverage a shared memory, this would alleviate the need for transfer completely.
However, only up to the point where the mapped qubit Hamiltonian needs to be transferred to the QPU.
Until we have a direct high-bandwidth connection between the CPU and QPU, this transfer of data
(also known as the \emph{data loading problem} in quantum computing) will remain the rate-limiting factor.
Therefore, within the scope of this more general problem, we deem the current implementation and design as scalable.
Furthermore, future improvements to aid in the transfer of data to the QPU that will be implemented into the
quantum stack will be accessible directly to the end-users of our integration because it only serves as a
middle-man between the two codes.

\section{Results and discussion}\label{sec:results}
To test our implementation we have studied the optical properties of the $F^0$-center
(neutral oxygen vacancy) in magnesium oxide, whose nature remains unclear despite the
many experimental~\cite{Chen1969,Kappers1970,Summers1983,Rosenblatt1989} and computational~\cite{Sousa2001,Rinke2012,Ertekin2013,Strand2019,Gallo2021,Lau2023,Vorwerk2023,Verma2023}
studies carried out in the past decades.
This data will allow us to compare the accuracy of the periodic \gls{rsdft} embedding with respect
to both experimental spectra as well as state-of-the-art \textit{ab initio} methods.

\subsection{Computational details}
\begin{figure*}
    \centering
    \includegraphics[width=0.85\textwidth]{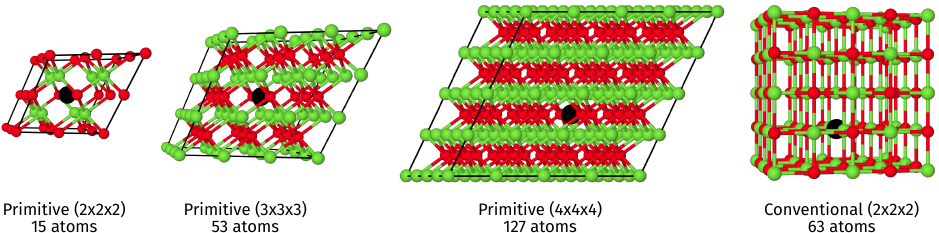}
    \caption{The four supercells used in the calculations. Magnesium atoms are in green,
    oxygen atoms are in red, the oxygen vacancy is colored in black.}
    \label{fig:MgO_supercell}
\end{figure*}

We have considered four different supercell sizes: the 2x2x2, 3x3x3 and 4x4x4 supercells constructed from
the primitive unit cell, and the 2x2x2 supercell constructed from the conventional unit cell; see
\cref{fig:MgO_supercell}.
For each system, we have first optimized the supercell parameters and geometry of the pristine system
with spin-unpolarized \gls{ks}-\gls{dft} and the \gls{pbe} functional~\cite{Perdew1996}, together with the
geometrical response valence triple-$\zeta$ basis set \cite{Hutter2024} and
correlation-consistent polarization functions (ccGRB-T basis set in CP2K). The core electrons were
described by the Goedecker-Teter-Hutter pseudopotential optimized for the \gls{pbe} functional~\cite{Goedecker1996,Hartwigsen1998,Krack2005}.
The \gls{dft} calculations were performed within the \gls{gpw} approach~\cite{Lippert1997,VandeVondele2005},
using plane-wave absolute and relative cutoffs of \num{1000} and \num{50} Rydberg, respectively,
and a 4-layer grid for the numerical integration.

To create the vacancy, we removed a single oxygen atom from each optimized pristine supercell and
relaxed again the geometry of all systems with the same settings as just discussed, while keeping
the cell parameters fixed.
Note that the ground state of the $F^0$-center in MgO is closed-shell, and
therefore spin-unpolarized \gls{dft} describes it well.
To perform the \gls{rsdft} embedding calculations we reduced the basis set to a double-$\zeta$
plus polarization (ccGRB-D) for all atoms but the six magnesium ones surrounding the vacancy,
for which we kept the triple-$\zeta$ one. In addition, we have also added the triple-$\zeta$ basis
functions of oxygen at the vacancy site, caculated as the center of mass of the six coordinated magnesium
atoms. Furthermore, we changed the pseudopotential to the one optimized for hybrid functionals.
We used the \gls{lda} functional in its \gls{sr} form~\cite{Perdew1992,Gill1996,Paziani2006} in
combination with a truncated Coulomb potential \cite{Guidon2009} for the \gls{lr} \gls{hf} component
of the functional, using a truncation radius of \qty{4.25}{\angstrom}.
The \gls{rs} parameter was set to \qty{0.14}{\bohr^{-1}}, which was optimally tuned by matching the
bandgap of the largest pristine supercell to the experimental bandgap value of
\qty{7.77}{\electronvolt} \cite{Roessler1967}.

\begin{figure}
    \centering
    \includegraphics[width=0.25\textwidth]{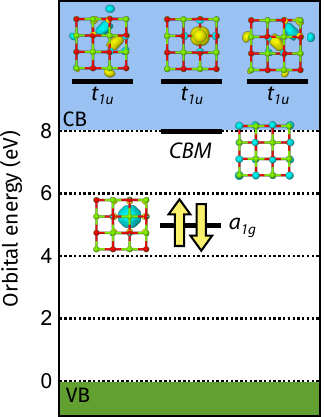}
    \caption{Band diagram of the $F^0$-center in magnesium oxide.
    The ground state is a closed-shell \ce{A_{1g}} singlet, with two electrons occupying
    a defect orbital within the gap. Singlet and triplet excitons occur when an electron
    from the mid-gap orbital is excited either to the fully delocalized \gls{cbm} state,
    or to one of the three \ce{t_{1u}} defect orbitals within the conduction band.
    The 4 defect orbitals, along with the \gls{cbm} one are included in the active space.}
    \label{fig:MgO_bands_diagram}
\end{figure}

The embedding calculations were performed by including 2 electrons and 5 orbitals (10 spin-orbitals in
total) in the \gls{as}, which are shown within the band structure diagram in \cref{fig:MgO_bands_diagram}.
Four of the five active orbitals are localized at the vacancy and are labeled according to the (localized)
octahedral symmetry around the defect (\ce{O_h} point group). The remaining orbital included corresponds
to the \gls{cbm} and is a fully delocalized conduction \ce{s}-band, which we label as \gls{cbm} in the
following.
The ground state energy of the embedded fragment Hamiltonian was obtained using the \gls{vqe} algorithm with
a \gls{uccsd} ansatz, and the resulting quantum circuits were simulated without the addition of artificial noise using Qiskit (version 0.45).

The absorption and emission spectra were obtained by calculating the lowest ten excitonic states
of singlet and triplet spin symmetry, respectively, using the \gls{qeom} approach~\cite{Ollitrault2020}.
The energies for the photoluminescence spectrum were computed at the relaxed geometry of the
\ce{^3T_{1u}} excited state, optimized using \gls{tddft} within the Tamm--Dancoff approximation~\cite{Hirata1999,Hehn2022} on top of the \ce{^1A_{1g}} ground state. The other computational settings
were the same as those used for the ground state geometry relaxation.
The \gls{as} embedding calculations for the singlet and triplet states were based on spin-unpolarized and
spin-polarized \gls{dft}, respectively, and were converged to an energy change of \qty{1e-8}{\hartree}.
At last, notice that the \gls{uccsd} ansatz for 2 electrons is equivalent to an exact diagonalization,
which we confirm by performing all calculations using the classical \gls{fci} solver of PySCF~\cite{Sun2020}
instead of \gls{vqe} plus \gls{qeom}.
Additional information on the computational details is provided in the \gls{si}.

\subsection{Absorption and photoluminescence spectra}
\begin{figure}
    \centering
    \includegraphics[width=\linewidth]{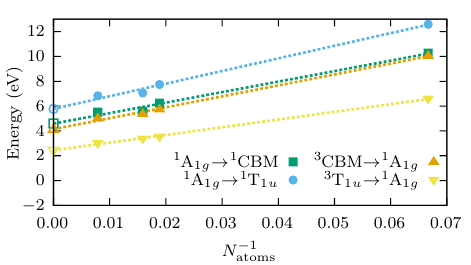}%
    \caption{Vertical absorption and emission energies as a function of the (inverse)
    number of atoms in the supercell. Filled green squares and blue circles correspond
    to calculated singlet absorption energies, while filled upside orange and downside
    yellow triangles correspond to calculated triplet emission energies. Dashed
    lines are linear extrapolation curves to the \gls{tdl}, whose value
    is marked with a corresponding empty symbol.}
    \label{fig:MgO_energy_extrap}
\end{figure}

A neutral oxygen vacancy in MgO, typically called $F^0$-center or \ce{V^0_O} vacancy, introduces a
\ce{1s}-type localized defect orbital at mid-gap that is doubly occupied in the \ce{^1A_{1g}}
electronic ground state. We shall call this orbital the mid-gap orbital. Three more defect-localized
degenerate one-particle states of \ce{t_{1u}} symmetry (\ce{p}-like orbitals) appear within the
conduction band. These three orbitals are energetically slightly above the \gls{cbm}, which
corresponds to the delocalized \ce{s}-band. The four orbitals localized at the oxygen vacancy and
the \gls{cbm} one are shown in \cref{fig:MgO_bands_diagram} and are believed to
be responsible for the optical properties of defective MgO; for this reason we included them in the
active space of our embedding calculation, along with the 2 electrons occupying the mid-gap defect
state.
\paragraph*{Absorption.}
Experimentally, it is well established that the absorption peak of the $F^0$-center is at
\qty{5.03}{\electronvolt}, which is extremely close to that of the $F^+$-center (that is, a
positively charged oxygen vacancy) at \qty{4.96}{\electronvolt} \cite{Chen1969,Kappers1970}.
We identified the vertical excitation energies corresponding to the transition of one electron
from the mid-gap orbital to either the \gls{cbm} or the \ce{t_{1u}} orbitals, denoted as
\ce{^1A_{1g}}$\to$\ce{^1CBM} and \ce{^1A_{1g}}$\to$\ce{^1T_{1u}}, respectively. In \cref{fig:MgO_energy_extrap}
we show in blue and green the energies obtained for these states as a function of the number of atoms
in the supercell. Notice that the energies calculated with our embedding method are obtained directly
from many-body wave functions, thus they account for electron-hole interactions and the exciton binding
energy.
We performed an extrapolation to the \gls{tdl} assuming a $N_\text{atoms}^{-1}$ convergence to
correct for finite-size effects, which is shown as dashed lines in \cref{fig:MgO_energy_extrap}.
Our best estimates are reported at the bottom of \cref{tab:MgO_singlet}, along with all
the energies obtained for the different supercell sizes.
\begin{table}
    \centering
    \begin{tabular}{cccc}
        \toprule
        cell & $N_\text{atoms}^{-1}$ & \ce{^1A_{1g}}$\to$\ce{^1CBM} & \ce{^1A_{1g}}$\to$\ce{^1T_{1u}} \\
        \midrule
        p 2x2x2      & \num{0.067} & \num{10.27} & \num{12.59} \\
        p 3x3x3      & \num{0.019} &  \num{6.24} &  \num{7.74} \\
        c 2x2x2      & \num{0.016} &  \num{5.59} &  \num{7.06} \\
        p 4x4x4      & \num{0.008} &  \num{5.52} &  \num{6.83} \\
        $\to \infty$ & $\to 0.0$   &  \num{4.60} &  \num{5.78} \\
        \bottomrule
    \end{tabular}
    \caption{Optical singlet absorption energies (in eV) of the $F^0$-center in MgO.
    The bottom row contains the values extrapolated to the \gls{tdl}.}
    \label{tab:MgO_singlet}
\end{table}
The transition to the \ce{^1CBM} state is predicted to be \qty{4.60}{\electronvolt}, while
for the (triply degenerate) one to the \ce{^1T_{1u}} state we get \qty{5.78}{\electronvolt}.
The experimental absorption peak at \qty{5.03}{\electronvolt} corresponds to the
\ce{^1A_{1g}}$\to$\ce{^1T_{1u}}, as the excitation to the \gls{cbm} is dipole-forbidden; we therefore
overestimate the experimental value by \qty{0.75}{\electronvolt}.
This result can be compared to a number of other computational studies on the $F^0$-center
in magnesium oxide, which have been performed with a large variety of methods from both the
solid-state physics and quantum chemistry communities. From \cref{tab:MgO_comparison} we can
see that most of the quantum chemistry methods overestimate the absorption to the localized
\ce{^1T_{1u}} state, while FN-DMC~\cite{Ertekin2013} and \ce{G_0W_0}-BSE~\cite{Rinke2012} reproduce
the experimental absorption peak, despite the fact that they likely target the \gls{cbm} state rather
than the defect one.
This may suggest that the transition to the \ce{^1T_{1u}} obtained with those method
is possibly overestimating the experimental absorption maximum.
Our calculations consistently predict a blue-shifted absorption by \qty{\sim 0.5}{\electronvolt}
compared to the embedded-BSE@DDH of~\citet{Vorwerk2023}.
One possible reason could be the small active space of 2 electrons in 5 orbitals for both
the ground state \gls{vqe}-\gls{uccsd} calculation and the subsequent \gls{qeom} one,
which does not allow sufficient relaxation of the \ce{t_{1u}} orbitals, as these
are empty in the ground state calculation and thus not optimized.
Possible solutions for this problem are discussed in Section\@~\ref{sec:conclusion}.
Interestingly, density-matrix embedding theory~\cite{Verma2023} predicts the absorption to the
\gls{cbm} state to be higher than the one to the localized state, perhaps owing to the missing
extrapolation to the \gls{tdl}.
The only method that underestimates the absorption is TDDFT, with an absorption peak centered
at \qty{4.85}{\electronvolt}~\cite{Strand2019}.
\begin{table*}
    \centering
    \begin{tabular}{lcccc}
        \toprule
        method           & structure & \ce{^1A_{1g}}$\to$\ce{^1CBM} & \ce{^1A_{1g}}$\to$\ce{^1T_{1u}} & emission\\
        \midrule
        qEOM-srLDA (this work)                & periodic & \num{4.60} & \num{5.78}  & \num{2.44} \\
        NEVPT2-DMET@ROHF   \cite{Verma2023}   & periodic &
        \num{5.67}\footnote{Value corresponding to the p 4x4x4 system.} & \num{5.24} & \num{2.89} \\
        embedded-BSE@DDH   \cite{Vorwerk2023} & periodic & \num{4.13} & \num{5.23}  &
        \num{2.93}\footnote{Value obtained for the emission from a different state than \ce{^3T_{1u}}.\label{fn:emission}} \\
        EOM-CCSD (GTOs)    \cite{Lau2023}     & periodic & -          & \num{5.31}  & - \\
        EOM-CCSD (PWs)     \cite{Gallo2021}   & periodic & -          & \num{5.28}  & \num{3.66} \\
        TDDFT@PBE0         \cite{Strand2019}  & periodic & -          & \num{4.85}  & \num{2.90} \\
        FN-DMC@PBE         \cite{Ertekin2013} & periodic & \num{5.00} & -           & \num{3.80}\textsuperscript{\ref{fn:emission}} \\
        \ce{G_0W_0}@LDA0-BSE \cite{Rinke2012} & periodic & \num{4.95} & -           & \num{3.40}\textsuperscript{\ref{fn:emission}} \\
        CASPT2(2,2)        \cite{Sousa2001}   & cluster  & -          & \num{5.44}  & \num{4.09} \\
        \midrule
        experiment & -        & dark       & \num{5.03} \cite{Chen1969,Kappers1970}  & \num{2.30} \cite{Summers1983,Rosenblatt1989} \\
        \bottomrule
    \end{tabular}
    \caption{Comparison of predicted optical absorption and photoluminescence energies (in eV)
    obtained with different computational methods. Different computational studies computed the
    emission energies from different states, see the footnotes and the main text for more information.}
    \label{tab:MgO_comparison}
\end{table*}

\paragraph*{Photoluminescence.}
The \gls{pl} of defective MgO is considerably more complicated than the absorption, and several
interpretations have been brought forward throughout the years.
The main source of ambiguity is the vicinity of the absorption peaks of the $F^0$,
$F^+$ and $F^{2+}$ centers, which are all likely to be excited by incoming irradiation at
\qty{5}{\electronvolt}, significantly complicating the assignment of the emission bands to the
correct point defect and electronic state.
Experimentally, there are two very distinct peaks visible in the \gls{pl} spectrum, one at
\qty{2.3}{\electronvolt} and one at \qty{3.2}{\electronvolt}~\cite{Summers1983,Rosenblatt1989}.
The former has been associated to an emission from the $F^0$-center, while the latter
to an emission from the $F^+$-center~\cite{Summers1983,Rosenblatt1989}.
The initial explanation for the long-lived nature of the \qty{2.3}{\electronvolt} band was
based on temperature-dependent experiments carried out on samples prepared in different ways
and containing different concentrations of $F^0$-centers and hydrogen impurities.
The proposed mechanism involves the escape of an electron in the conduction band upon
excitation of the $F^0$-center, leaving behind a positively charged oxygen vacancy, an $F^+$-center.
Hydrogen impurities in the sample then act as traps for the mobile electrons, which may be thermally
released back at a later time into the conduction band. Two processes can happen with the released
electrons:
they may encounter $F^+$-centers left behind after the absorption process, in which case they recombine
in an excited \ce{^1T_{1u}} state of the $F^0$-center that quickly emits light, or they may be
recaptured by \ce{H^-} traps, slowing down the overall emission process.
An alternative, more straightforward interpretation is that the emission band is simply due to
triplet phosphorescence from a localized \ce{^3T_{1u}} state at the defect, accessed via
inter-system crossing from the excited \ce{^1T_{1u}} state.
This second interpretation is the most accepted explanation in recent computational studies,
corroborated by calculations based on advanced \textit{ab initio}
methodologies~\cite{Strand2019,Gallo2021,Vorwerk2023,Verma2023}
(in contrast to the first interpretation based
on older semi-empirical methods and experiments~\cite{Summers1983,Rosenblatt1989}).

In light of this analysis, we also investigate the second pathway as the leading emission process
and compute the photoluminescence from the relaxed \ce{^3T_{1u}} structure and state.
Our results are shown by the orange and yellow curves in \cref{fig:MgO_energy_extrap}
and listed in \cref{tab:MgO_triplet}.
\begin{table}
    \centering
    \begin{tabular}{cccc}
        \toprule
        cell & $N_\text{atoms}^{-1}$ & \ce{^3T_{1u}}$\to$\ce{^1A_{1g}} & \ce{^3CBM}$\to$\ce{^1A_{1g}} \\
        \midrule
        p 2x2x2      & \num{0.067} & \num{6.59} & \num{10.05} \\
        p 3x3x3      & \num{0.019} & \num{3.54} &  \num{5.74} \\
        c 2x2x2      & \num{0.016} & \num{3.38} &  \num{5.38} \\
        p 4x4x4      & \num{0.008} & \num{3.02} &  \num{5.02} \\
        $\to \infty$ & $\to 0.0$   & \num{2.44} &  \num{4.14} \\
        \bottomrule
    \end{tabular}
    \caption{Photoluminescence energies (in eV) of the $F^0$-center in MgO.
    The bottom row contains the values extrapolated to the \gls{tdl}.}
    \label{tab:MgO_triplet}
\end{table}
The predicted value for the emission from the localized triplet state is
\qty{2.44}{\electronvolt} and is in very good agreement with the experimental
value of \qty{2.3}{\electronvolt}. In particular, we are much closer  to this value
than other methodologies, which consistently predict higher energies for this emission
band as shown in the last column of \cref{tab:MgO_comparison}.
Nevertheless, one has to be careful when comparing different theoretical works, since these
focused on different states or mechanisms.
The work by~\citet{Vorwerk2023} reported \qty{2.93}{\electronvolt} as the \gls{pl} from
the $F^+$-center, hence to be compared with the experimental value of \qty{3.2}{\electronvolt}.
The earlier work based on FN-DMC~\cite{Ertekin2013} and \ce{G_0W_0}-BSE~\cite{Rinke2012} reported
the emission from the singlet state (whether the \gls{cbm} or \ce{T_{1u}} state is unclear),
and have concluded that the assignment from the experimental studies should be re-evaluated,
with the \qty{3.2}{\electronvolt} peak assigned to the $F^0$-center rather than the $F^+$ center.
The methodologies based on quantum chemistry methods, that is NEVPT2-DMET, EOM-CCSD, TDDFT and CASPT2,
calculate the transition \ce{^3T_{1u}}$\to$\ce{^1A_{1g}} like us and compare their results against
the \qty{2.3}{\electronvolt} band~\cite{Sousa2001,Strand2019,Gallo2021,Verma2023}.
Here we do the same and we get the best theoretical result so far, which corroborates the experimental
assignment of the \qty{2.3}{\electronvolt} band to the $F^0$-center, though, originating from a
different mechanism than the originally proposed one.

\section{Conclusions}\label{sec:conclusion}
We developed a general framework for hybrid quantum-classical molecular
and periodic embedding calculations based on an orbital space separation of the system into fragment
and environment.
This framework has been implemented in the CP2K package, leveraging many of its functionalities and
taking advantage of its high parallel efficiency.
The modular nature of the implementation allows to easily develop several types of embedding schemes
and to couple different solvers to obtain ground and excited states energies and properties
of the embedded fragment Hamiltonian. It supports both classical wave function and quantum circuits
ansatzes, and the communication between CP2K and the solver is handled by sockets, which
seamlessly integrates within current supercomputing facilities, but is also ready for a more
quantum-centric high-performance computing vision.

To demonstrate the potential of the new framework in practice, we have implemented a range-separated DFT
embedding scheme that enables the study of both finite and extended systems. This approach is
essentially an extension of multiconfigurational range-separated DFT to periodic boundary conditions
and relies on the range-separation of the two-electron integrals in long- and short-range components.
Within this approach, any correlated wave function method can in principle be coupled with \gls{dft}
in a self-consistent scheme, whereby the former is used to obtain the spectrum of a fragment Hamiltonian,
and the latter to construct an embedding potential generated by the environment degrees of freedom.
In particular, as part of this work, we have implemented an interface to Qiskit Nature~\cite{QiskitNatureCP2K}
that allows to map the fermionic fragment Hamiltonian to a qubit Hamiltonian, whose ground and excited
states can be obtained with the quantum algorithm of choice.

The developed \gls{rsdft} embedding scheme has a wide application scope, allowing the investigation
of both strongly correlated molecular systems, as well as localized electronic states in materials,
such as those arising from vacancies and impurities.
To this end, we have demonstrated its accuracy and applicability by studying the optical properties
of the neutral oxygen vacancy in MgO, whereby both defect-localized and delocalized states have been
treated on equal footing, and the low-lying spectrum of the embedded fragment Hamiltonian has been
calculated by \gls{vqe} and \gls{qeom}, in combination with the \gls{uccsd} ansatz.
Our calculations for the absorption spectrum predict a peak at \qty{5.78}{\electronvolt}, a value
that overestimates the experimental result by \qty{0.75}{\electronvolt}, but which lies in the same ballpark
as other sophisticated computational approaches.
On the other hand, the predicted \gls{pl} emission of \qty{2.44}{\electronvolt} from the \ce{^3T_{1u}}
state almost perfectly matches with the experimentally measured signal at \qty{2.3}{\electronvolt},
and provides new insight on a system that has eluded state-of-the-art \textit{ab initio}
approaches for the last decade.
While the accuracy of the method for the absorption leaves room for improvement, and the excellent
agreement for the emission is certainly helped by favorable error compensation, the present study
shows that current hybrid quantum-classical algorithms can compete with classical state-of-the-art
\textit{ab initio} methodologies for problems beyond simple model systems.

Many possible future directions are envisioned based on this work. On one hand, the periodic
\gls{rsdft} embedding scheme can be extended in several ways. For instance, introducing
orbital optimization would allow to incorporate the feedback from the correlated active space
wave function onto the inactive long-range component; this would be particularly important for
accounting the changes in the environment when targeting states other than the ground state.
Furthermore, it would make the embedding scheme truly variational, significantly simplifying
the calculation of analytical forces.
State-averaging would also be a useful extension, allowing a more balanced description of several
states simultaneously. This is fundamental in cases where near-degeneracies are prominent, such
as in molecules and materials containing open-shell transition metals.
While optimally-tuned short-range \gls{lda} performed well in this study, implementing more
\gls{sr} functionals will provide alternatives to cases where \gls{lda} is not sufficiently
accurate.

Future directions that are not strictly tied to the \gls{rsdft} embedding
scheme are also envisaged. For instance, one possibility is to implement orbital localization
schemes based on maximally localized Wannier functions and pair natural orbitals,
which would allow to study pristine solid-state materials, where localized states do not arise
naturally due to symmetry-breaking of the supercell.
Owing to the local nature of electron correlation, orbital localization could also simplify the
construction of hardware efficient ansatzes, thereby potentially increasing the maximum size
of the active space.
Finally, CP2K can be interfaced to other classical active space solvers, such as those based on
selected configuration interaction, which would also allow to study larger fragment Hamiltonians
solely on classical hardware.

\section*{Acknowledgements}
The authors thank Valery Weber for insightful discussions and guidance during the early stages of the code development.
This research was supported by the NCCR MARVEL, a National Centre of Competence in Research, funded
by the Swiss National Science Foundation (grant number 205602). This work has been also supported by
the Swiss National Science Foundation in the form of Ambizione grant No. PZ00P2\_174227 (VVR).
IBM, the IBM logo, and ibm.com are trademarks of International Business Machines Corp., registered
in many jurisdictions worldwide. Other product and service names might be trademarks of IBM or other
companies. The current list of IBM trademarks is available at \url{https://www.ibm.com/legal/copytrade}.

\section*{Supplementary Information}
The Supplementary Information contains the derivation of the \gls{hf} approach within the \gls{gpw}
formalism, showing the expression of the potentials expressed in the molecular orbital basis.
It also contains the technical details about the message passing interface used to couple CP2K
and Qiskit nature, and additional computational details regarding the calculations performed.
All input and output files, structure files, tabulated raw data and scripts to perform the simulations
are available on the Materials Cloud platform at
\href{https://doi.org/10.24435/materialscloud:47-6g}{10.24435/materialscloud:47-6g}.

\bibliography{references}

\end{document}


\title{Supplementary Information: A general framework for active space embedding methods: applications in quantum computing}

\author{Stefano Battaglia}
\email{stefano.battaglia@chem.uzh.ch}
\thanks{These authors contributed equally to this work.}
\affiliation{Department of Chemistry,
    University of Zurich,
    Winterthurerstrasse 190,
    CH-8057 Zürich,
    Switzerland
}

\author{Max Rossmannek}
\email{oss@zurich.ibm.com}
\thanks{These authors contributed equally to this work.}
\affiliation{Department of Chemistry,
    University of Zurich,
    Winterthurerstrasse 190,
    CH-8057 Zürich,
    Switzerland
}
\affiliation{IBM Quantum, IBM Research – Zurich, CH-8803 Rüschlikon, Switzerland}

\author{Vladimir V. Rybkin}
\altaffiliation[now at: ]{HQS Quantum Simulations GmbH,
    Rintheimer Strasse 23,
    DE-76131 Karlsruhe,
    Germany
}
\affiliation{Department of Chemistry,
    University of Zurich,
    Winterthurerstrasse 190,
    CH-8057 Zürich,
    Switzerland
}

\author{Ivano Tavernelli}
\affiliation{IBM Quantum, IBM Research – Zurich, CH-8803 Rüschlikon, Switzerland}

\author{Jürg Hutter}
\affiliation{Department of Chemistry,
    University of Zurich,
    Winterthurerstrasse 190,
    CH-8057 Zürich,
    Switzerland
}

\maketitle

\section{Theory}

\subsection{The Hartree-Fock energy in the GPW formalism}
We start by defining the periodic \gls{hf} energy expression in the \gls{gpw} formalism, which reads
\begin{equation}
     \label{eq:ehf_periodic}
     E^{\text{HF}} = \sum_{ij} h^0_{ij} D_{ij} + E_H[\rho_{\text{tot}}] + E_{\text{ovlp}}
                   - E_{\text{self}} + E_{\text{HFX}}[\rho] \, .
\end{equation}
Here, $h^0_{ij}$ are one-electron integrals, $E_H$ is the Hartree energy, $E_{\text{HFX}}$ is the \gls{hf}
exchange energy, and $E_{\text{ovlp}}$ and $E_{\text{self}}$ are the overlap and self-interaction energy,
respectively.
The labels $i$ and $j$ refer to molecular orbitals defined in terms of periodic Gaussian basis functions at the
$\Gamma$ point, $\rho = \rho(\mbf{r})$ is the electron density, and $\rho_\text{tot} = \rho_\text{tot}(\mbf{r})$
is the total charge density. The one-electron integrals, $h^0_{ij} = \braket{\psi_i|\hat{h}^0|\psi_j}$,
are defined in terms of the following one-body operator
\begin{equation}
    \label{eq:h_gpw}
    \hat{h} = \hat{T} + \hat{V}_{\text{loc}}(\mbf{r}) + \hat{V}_{\text{nl}}(\mbf{r},\mbf{r}') \, ,
\end{equation}
where $\hat{T}$ is the kinetic energy operator, $\hat{V}_{\text{loc}}(\mbf{r})$ is a local pseudopotential
term, and $\hat{V}_{\text{nl}}(\mbf{r},\mbf{r}')$ a non-local pseudopotential contribution (see
\cite{Lippert1997} and \cite{VandeVondele2005} for more details). Notice that compared to the
typical definition of the one-body operator in the molecular case, \Cref{eq:h_gpw} does not
include the core potential, since it is already included in the Hartree term, as discussed below.
The total charge density, $\rho_\text{tot}$, is obtained by the sum of the electron density $\rho$, given by
\begin{equation}
    \label{eq:rho}
    \rho = \rho(\mbf{r}) = \sum_{ij} D_{ij} \psi^*_i(\mbf{r}) \psi_j(\mbf{r}) \, ,
\end{equation}
and the nuclear charge distribution, which is expressed as a sum of Gaussian functions $n_P^c = n_P^c(\mbf{r})$
\begin{equation}
    \rho_{\text{tot}} = \rho_{\text{tot}}(\mathbf{r}) = \rho(\mathbf{r}) + \sum_P n_P^c(\mathbf{r}) \, .
\end{equation}
The total charge density is used in the evaluation of the Hartree energy, $E_H[\rho_{\text{tot}}]$, which accounts
for the classical Coulomb interaction. The self-interaction term, $E_{\text{self}}$, removes the self-interaction of
the nuclear charge distribution with itself (akin to the exact \gls{hf} exchange energy for the electronic component),
while the overlap energy, $E_{\text{ovlp}}$, accounts for the overlap between the core and electron densities.
The last term on the right-hand side of \cref{eq:ehf_periodic} is the \gls{hf} exchange energy,
$E_{\text{HFX}}[\rho]$.

To derive the working equations for an active space embedding approach using the \gls{gpw} formalism, it is convenient
to formally express the right-hand-side terms of \Cref{eq:ehf_periodic} as a function of the \gls{1rdm} and the Hartree
and exact exchange potentials in the molecular orbital basis as
\begin{align}
    \label{eq:E_H}
    E_H[\rho_{\text{tot}}] + E_{\text{ovlp}} -  E_{\text{self}} &=
    \sum_{pq} V^{\text{core}}_{ij} D_{ij} + \overbrace{\frac{1}{2}\sum_{ij} V^H_{ij}[\rho]D_{ij}}^{= E_H[\rho]}
    + \sum_{PQ} V^{nn}_{PQ} \, ,
\end{align}
and
\begin{align}
    \label{eq:E_HFX}
    E_{\text{HFX}}[\rho] &= -\frac{1}{4} \sum_{ij} V^{\text{HFX}}_{ij}[\rho] D_{ij} \, ,
\end{align}
with the potentials\footnote{Notice that here and in the remainder of the manuscript we switch the dependence
between the electron density and the \gls{1rdm} depending on the context, keeping in mind their relationship.}
\begin{align}
    \label{eq:V_H}
    V^{H}_{ij}[\rho] &= 
    \sum_{kl} D_{kl} g_{ijkl} \, , \\
    \label{eq:V_HFX}
    V^{\text{HFX}}_{ij}[\rho] &= \sum_{kl} D_{kl} g_{ikjl} \, .
\end{align}
In \Cref{eq:E_H} we have formally isolated the energy contributions due to the nuclear-nuclear interaction,
$V^{nn}_{PQ}$, and the electron-nuclear interaction through the core potential, $V^{\text{core}}_{ij}$.
The classical electronic Hartree and HF exchange potentials are expressed in terms of the
two-electron integrals, $g_{ijkl}$.

\section{Implementation}

\subsection{Message passing}

Whether a UNIX or INET socket is used for the communication between CP2K and Qiskit Nature does not affect
the messages that are being sent. In either case, both programs will be given read and write access
to the bi-directional socket file allowing them to pass binary data back and forth.
We have agreed upon 8 message types which are distinguished by a header of 12 characters in length.
The different messages are used to synchronize the processes and indicate their readiness to one another
as well as to announce and subsequently share the actual payloads of information.
In what follows, we briefly summarize the different message types.

The \texttt{STATUS} message is the first one being sent from CP2K to Qiskit Nature after the socket has been
successfully established.
During the first embedding iteration the only valid response from Qiskit Nature is the \texttt{READY} message,
indicating that CP2K may start sending the active space Hamiltonian.
CP2K will proceed to do so via the \texttt{TWOBODY} message followed by the number of spins, the number
of active molecular orbitals, the number of active electrons, and the multiplicity.
All of these variables should be transferred as 32-bit integers.
In immediate succession, CP2K should transfer the active \glspl{eri}.
First, the alpha-alpha spin intgegrals and then, if the number of spins is greater than 1,
the alpha-beta spin integrals, beta-beta spin integrals, and the alpha-beta overlap matrix.
All of this data only has to be transferred once, since its values are not going to change throughout
the course of the self-consistent embedding.
Qiskit Nature will acknowledge the payload with a \texttt{RECEIVED} message.

Next, CP2K will transfer the one-body terms of the active space Hamiltonian.
This occurs after another round of \texttt{STATUS}/\texttt{READY} messages, followed by the \texttt{ONEBODY}
header and subsequent payload of the alpha-spin and optionally beta-spin integrals.
Upon completion, Qiskit Nature will not acknlowedge the transfer and instead immediately proceed with finding the
ground state solution to this active space Hamiltonian.
It will announce its results to CP2K using the \texttt{HAVEDATA} message.

When CP2K receives this message, it will query the updated \gls{1rdm} from Qiskit Nature using the
\texttt{GETDENSITY} request.
This will transfer the alpha-spin and optionally beta-spin density matrices and initiates the next
embedding iteration.
Since the two-body integrals do not require updating, CP2K will proceed immediately with the updated
one-body integrals as described above.

Once the embedding algorithm has converged, CP2K should inform Qiskit Nature using the \texttt{QUIT} message.

\section{Computational details}
As reported in the main text, the \gls{dft} calculations were performed within the \gls{gpw} formalism.
In practice, this means that the classical electrostatic potential is computed within \gls{gpw},
but the \gls{hf} exchange is computed using analytical integrals in real space.
The active space integrals are instead all computed within the \gls{gpw} approach, using the same
cutoffs and number of grids as for the \gls{dft} part. This means absolute and relative cutoffs of
\num{1000} and \num{50} Rydberg, respectively, and 4 grids.
As one can see from equation 18 of the main text, the embedding potential is computed
indirectly by subtracting from the total Fock matrix built the density at iteration $i$, i.e. $\rho^{(i)}$,
the active components of the long-range Hartree and exact exchange potentials. These latter two are both
evaluated using the active space integrals, while the total Fock matrix using a combination of \gls{gpw}
and analytical integrals. It is therefore important to use sufficiently large values for the \gls{gpw}
cutoffs, such that no significant difference is observed with respect to the \gls{hf} exchange integrals
evaluated analytically during Fock matrix build.

The \ce{^3T_{1u}} electronic state is the lowest triplet state for all supercells but the smallest, at
least within spin-polarized \gls{dft} and the computational settings used in this work.
To ensure a consistent set of optimized orbitals across all supercells and be able to perform the energy
extrapolation to the \gls{tdl}, we have used the maximum overlap method \cite{Barca2018} to converge the SCF
calculation of the primitive 2x2x2 supercell to the localized \ce{^3T_{1u}} state.

For all calculations we have used the CP2K default thresholds, unless specified differently in the main
text, and for the convergence of the SCF iterations, for which we have tighten the threshold from the
default \num{10e-5} to \num{10e-6}.

\bibliography{references}